\documentclass[12pt]{article}
\usepackage{latexsym}
\usepackage{amssymb}
\usepackage{amsthm}
\usepackage{epsfig}
\usepackage{graphicx}
\usepackage{placeins}
\graphicspath{{./}{Figures/}}
\usepackage{mathtools} 
\usepackage{enumitem}
\usepackage{xcolor}
\usepackage{easyReview}
\usepackage{algorithm}
\usepackage[noend]{algpseudocode}

\usepackage{hyperref}
\hypersetup{
    linkcolor=blue,  
    linkbordercolor=red,
    linktoc=all,     
    linktocpage
}

\def\dsp{\def\baselinestretch{1.25}\large\normalsize}
\dsp
\allowdisplaybreaks

\begin{document}
\title{Analytical solution of electromagnetic heating in a stratified absorbing medium}
\author{Ulises Jaime-Yepez \; and \;  
Hongyun Wang\footnote{Corresponding author, hongwang@ucsc.edu} 
\\
Department of Applied Mathematics \\
University of California, Santa Cruz, CA 95064, USA\\
}
\maketitle

\begin{abstract}
We consider the temperature evolution in a skin tissue when it is exposed to an electromagnetic wave. The heating is caused by the absorbed power as the electromagnetic wave attenuates inside the skin. In many applications, the characteristic depth of attenuation is small in comparison with the lateral size of the electromagnetic beam. Because of this separation of length scales, the lateral heat conduction is negligible at leading order. Previously, we studied electromagnetic heating in a skin of a single homogeneous layer. We derived the leading-term analytical solution and the asymptotic corrections up to the second order for lateral heat conduction and for oblique incidence. In this study, we derive the analytical solution in the case of perpendicular incidence on a stratified skin tissue with a layer-specific profile for electromagnetic attenuation. It should be emphasized that in the current study, the heat capacity and conductivity are still uniform over the entire skin tissue; only the electromagnetic attenuation is stratified resulting in a layer-specific heating profile.  
\end{abstract}


\clearpage

%
\section{Introduction}
Millimeter-wave (MMW) systems are present in many civilian and defense applications. 
These systems utilize electromagnetic waves
of wavelength in millimeters corresponding to the frequency range of 30--300 GHz. 
The biological effects of MMW exposure on humans are mainly attributed to the 
heating of skin tissue caused by the absorbed electromagnetic power.  
The thermal effects of  electromagnetic absorption have been studied extensively. 
A particular thermal effect is the possible thermal burn damage on skin \cite{Parker_2024}.
A number of studies on thermal effects of MMW exposures were compiled in 
a special issue of the Journal of Directed Energy \cite{Whitmore_2021, Miller_2021, 
Cook_2021, Cook_2021B, Haeuser_2021, Cobb_2021, Parker_2021, Parker_2021B}. 
To establish safety standards on MMW human exposure, it is important to accurately 
model the spatial variation of electromagnetic energy deposition 
and the resulting temperature rise in realistic exposure scenarios \cite{Hirata_2021}.

In MMW exposures, the skin temperature increase is driven by the absorbed electromagnetic power 
during its propagation and attenuation in skin. 
The attenuation rate of an electromagnetic wave in skin determines the depth scale at which 
a $(1/e)$ fraction of the electromagnetic power survives and the remainder has been absorbed. 
For frequencies between 30 and 300 GHz, the electromagnetic penetration depth is well below millimeter
\cite{Parker_2017, Parker_2017B}. 
Accordingly, electromagnetic power in this frequency range is virtually absorbed within 
a few millimeters, producing a strong temperature gradient in the depth direction. 

In the lateral directions along the skin surface, heating varies with the power density 
distribution over the beam cross section, which spans a lateral scale of centimeters or larger
in applications. This separation of length scales between depth and lateral directions leads 
to a small depth-to-lateral scale ratio $\varepsilon$ in the system after nondimensionalization. 
The effect of heat conduction is inversely proportional to the square of the length scale. 
When $\varepsilon$ is small, the effect of heat conduction in lateral directions is small 
in comparison with the dominant heat conduction in the depth direction. 

Previously, we studied MMW exposures on skin consisting of a single layer with 
a uniform attenuation rate $\mu$. 
We derived the analytical solution of skin temperature distribution
in the case of $\varepsilon = 0 $ when the lateral heat conduction is neglected  \cite{WBZ_2020C}. 
In the presence of a small but finite depth-to-lateral scale ratio $\varepsilon $, 
we expanded the skin temperature distribution in powers of $\varepsilon$ and 
derived a two-term asymptotic solution \cite{Jaime-Yepez_2024}.
The two-term asymptotic approximation is accurate up to $O(\varepsilon^2) $ and 
captures the effect of lateral heat conduction even at a moderate $\varepsilon = 0.1$.
The analytical solution in the case of $\varepsilon = 0$ corresponds to the leading term 
in the two-term asymptotic solution for small $\varepsilon >0$. 

In this study, we investigate the effect of stratified electromagnetic power absorption 
in skin consisting of multiple layers with attenuation rates $\{\mu_j, \, j = 1, 2, \ldots \}$. 
In layer $j$, the electromagnetic power penetrating beyond depth $z$ decays exponentially 
with rate $\mu_j$. The density of heat source in layer $j$ is given by the negative derivative of 
the electromagnetic power surviving beyond depth $z$. 
We focus on the case of perpendicular incidence and $\varepsilon = 0$ 
(i.e., neglecting the effect of lateral heat conduction). 
We derive a closed form analytical solution for skin temperature evolution caused by 
stratified electromagnetic power absorption in a multi-layer skin. 
We need to clarify that in the current study we focus only on the stratified 
electromagnetic power absorption in skin. The heat capacity and conductivity 
are still assumed to be uniform in the entire skin tissue. 
The general case of a multi-layer skin with distinct heat conductions 
will be investigated in a subsequent study. 

One possible application of analytical solutions developed here is in inverse problems in which 
internal temperatures are inferred from measured surface temperatures 
using iterative optimization procedure which requires 
fast and reliable evaluation of skin temperature evolution \cite{WBZ_2023}. 
For a multi-layer skin, a closed-form analytical solution that is easy to compute 
also provides an opportunity of exploring the skin layer structure 
from measured surface temperatures when the incident beam power density is 
varied in time. 

This paper is organized as follows. Section 2 introduces the model of electromagnetic 
power absorption in a multi-layer skin and the resulting temperature evolution 
caused by the absorbed power. 
Section 3 non-dimensionalizes the temperature evolution model, 
using physical scales similar to those in \cite{Jaime-Yepez_2024, Wang_2025}. 
Subsequently, it writes the solution of the non-dimensional system as a combination 
of building blocks. Then all governing systems of building blocks are reduced to 
the parameter-free initial value problem (IVP) of the heat equation with 
the standard exponentially decaying heat source for $z > 0$. 
Section 4 applies Duhamel's principle to take care of the heat source in the parameter-free IVP
and derives a close-form analytical solution. 
Then it traces back the pathway of analytical reduction involved and 
construct the roadmap for evaluating the skin temperature distribution. 
It also numerically verifies the validity of the close-form analytical solution obtained. 
Section 5 summarizes the main results of the study. 


\section{Electromagnetic Absorption in a Multi-layer Skin and Temperature Evolution} \label{temp_evolu}
\subsection{Skin, incident beam and the coordinate system}
\begin{figure}[!h]
\vskip 0.0cm
\centering
\includegraphics[width=3.5in]{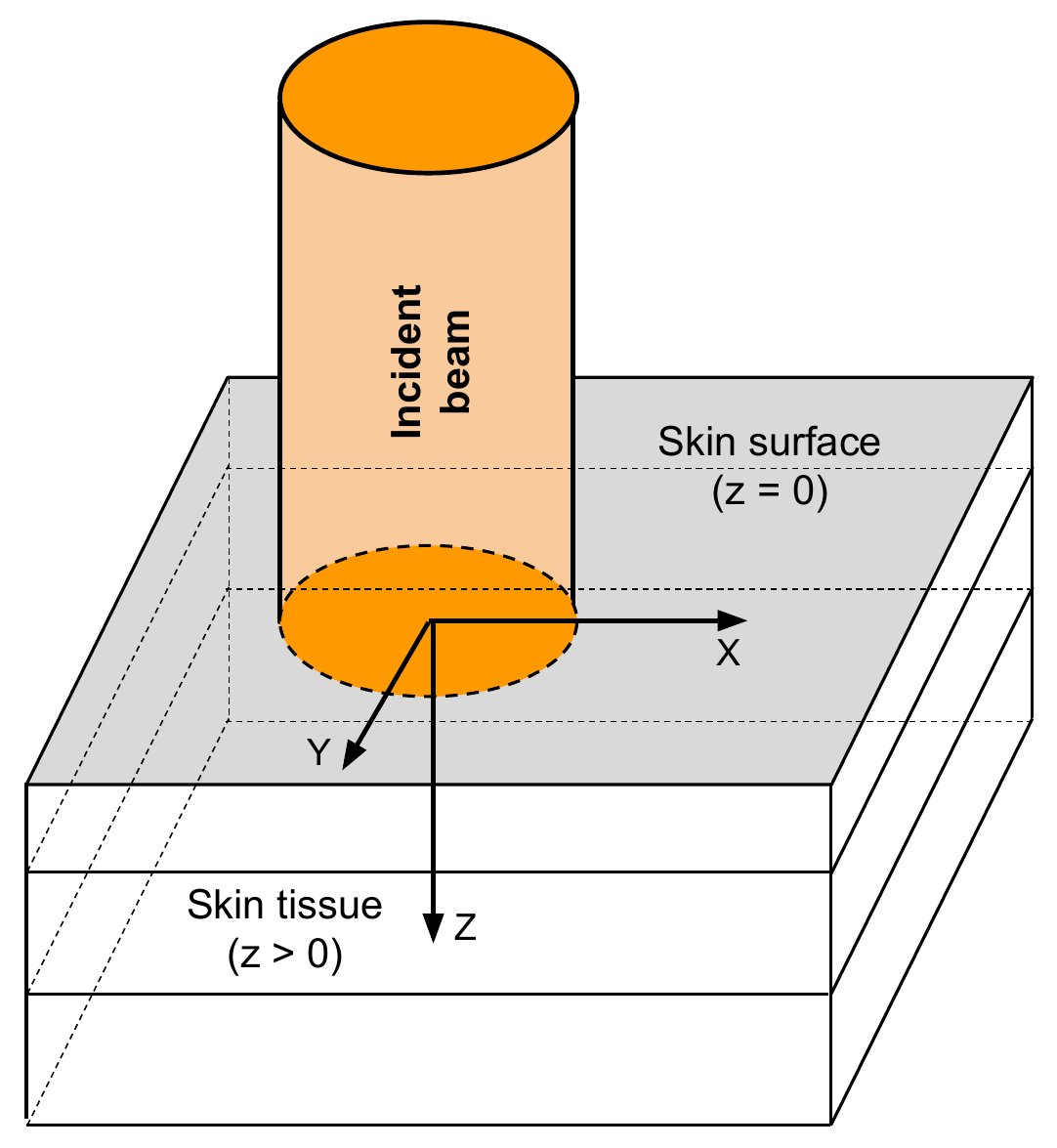} 
\vskip -0.0cm
\caption{A multi-layer skin, incident beam, and the coordinate system. }
\label{fig_01}
\end{figure}

We consider the situation where a skin tissue with a flat surface is exposed to 
an incident electromagnetic wave perpendicular to the skin surface. 
We establish the coordinate system as described in our previous studies
\cite{WBZ_2020C, Jaime-Yepez_2024}. 
In Figure \ref{fig_01}, the normal direction into the skin tissue is defined as the positive $z$-axis, 
with $z$-coordinate representing the depth into the skin and $z=0$ at the skin surface. 
The $(x, y)$-plane is the flat skin surface with, $(0, 0)$ selected as the center of the beam spot 
projected on the skin surface. 

\subsection{Stratified attenuation of electromagnetic power in skin}
Let $P_d^{\text{(i)}}(x, y) \equiv P_d^{(i)}\rho(x, y)$ be the incident beam power density 
arriving at the skin surface. 
$P_d^{(i)}$ is the beam center power density and $\rho(x, y)$ describes 
the relative power density distribution 
over the beam cross-section. By definition, is satisfies $\rho(0, 0) = 1$. 

Out of the power density arriving at the skin surface, $\alpha$ fraction passes through  
into the skin; the rest is reflected. 
The power density passing through beyond $z = 0$ is 
\begin{equation}
P_d^{(a)}(x, y)  = P_d^{(a)}\rho(x, y), \qquad P_d^{(a)} \equiv \alpha P_d^{(i)}
\label{Pd_xy}
\end{equation}
where $P_d^{(a)} \equiv \alpha P_d^{(i)}$ is the power density 
passing through beyond the skin surface at the beam center 
(and thus, is absorbed into the skin). 
Note that in general the transmission coefficient $\alpha$ varies with 
the electromagnetic frequency of the beam.  
In our mathematical formulation for solving the skin temperature evolution, we bundle 
the effect of the transmission coefficient $\alpha $ and 
the incident beam power density $P_d^{(i)}$ into one parameter 
$P_d^{(a)} \equiv \alpha P_d^{(i)}$. 

The electromagnetic thermal heating in skin is caused by the beam power absorbed 
per unit volume, which comes from the attenuation of electromagnetic power 
as it propagates in skin. 
Let $P_d^{(a)}(x, y, z)$ denote the power surviving through beyond depth $z$. 
Based on the Beer-Lambert Law \cite{Swinehart_1962}, $P_d^{(a)}(x, y, z)$ 
is governed by 
\begin{equation}
\begin{dcases}
\frac{\partial P_d^{(a)}(x, y, z)}{\partial z} = - \mu(z) P_d^{(a)}(x, y, z) \\[1ex]
P_d^{(a)}(x, y, z)\Big|_{z = 0} = P_d^{(a)}\rho(x, y)
\end{dcases}
\label{Beer-Lambert}
\end{equation}
where $\mu(z)$ is the attenuation rate (which varies with the depth). 
\begin{figure}[!h]
\vskip 0.0cm
\centering
\includegraphics[width=3.25in]{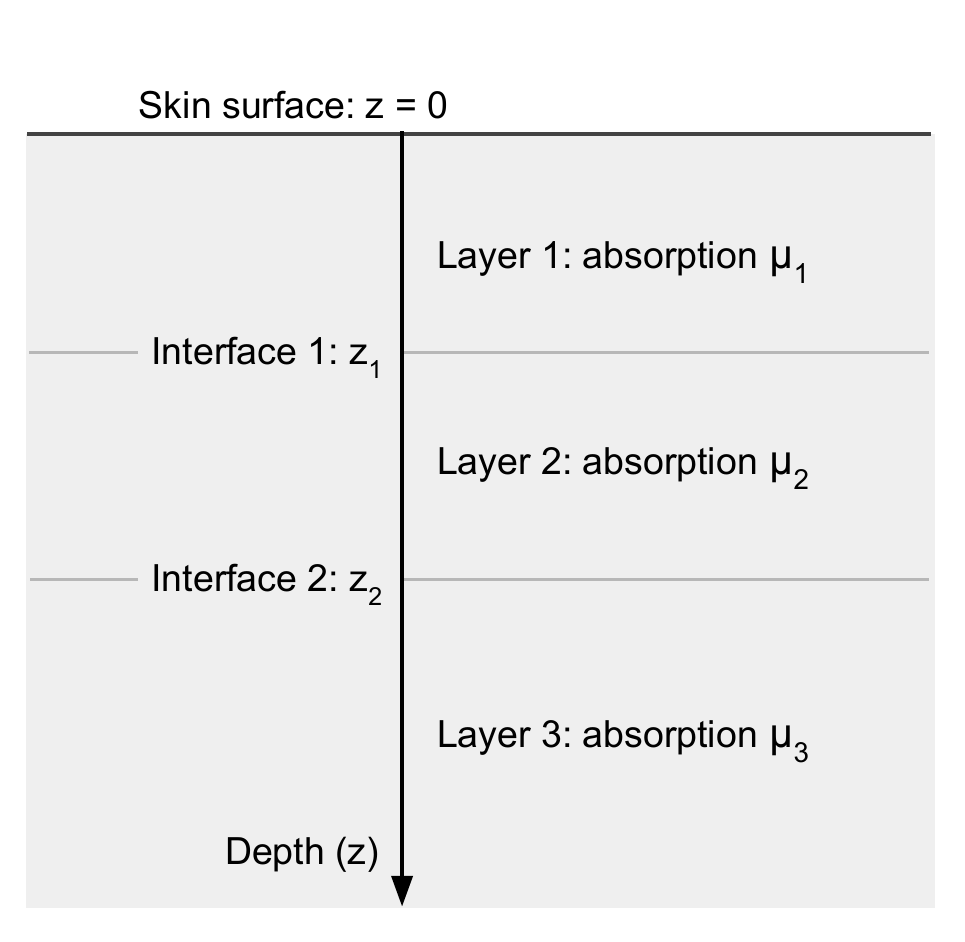} 
\vskip -0.0cm
\caption{A skin tissue of $M$ layers in depth. Layers are described by interfaces 
$\{(z_j)_{j=0}^M \}$. Layer $j$ has attenuation rate $\mu_j$. 
A case of $M = 3$ is illustrated.}
\label{fig_02}
\end{figure}

We consider a skin consisting of $M$ layers, with each layer 
having a constant attenuation rate $\mu$. 
As illustrated in Figure \ref{fig_02}, $\mu_j$ is the attenuation rate of layer $j$ and  
$z_j$ is the interface between layer $j$ and layer $(j+1)$.
For mathematical convenience, we view the air outside skin as layer 0. 
The air-skin interface is $z_0 = 0$, the last layer ends at $z_M = +\infty$, and 
the thickness of layer $j$ is $\Delta z_j \equiv z_j - z_{j-1}$, $\; j = 1, 2, \ldots, M$. 
\begin{align}
\text{Layer } j{:} \;\; & \mu(z) = \mu_j, \quad z \in (z_{j-1}, \; z_{j-1}{+}\Delta z_j), \quad j = 1, 2, \ldots, M
\label{mu_profile} \\[0ex]
& z_0 \equiv 0, \quad z_M \equiv +\infty, \quad \Delta z_j \equiv z_j - z_{j-1} 
\nonumber 
\end{align}
For stratified attenuation rate profile \eqref{mu_profile}, solution 
$P_d^{(a)}(x, y, z)$ of \eqref{Beer-Lambert} is 
\begin{align}
& P_d^{(a)}(x, y, z) = P_d^{(a)} \rho(x, y) S(z) \nonumber  \\[1ex] 
& S(z) = \begin{dcases}
e^{-\mu_1 z}, & z \in (0, \; \Delta z_1) \\[0ex] 
e^{-\mu_1 \Delta z_1} e^{-\mu_2 (z-z_1)}, & z \in (z_1, \; z_1{+}\Delta z_2) \\[0ex] 
e^{-(\mu_1 \Delta z_1+\mu_2 \Delta z_2)} e^{-\mu_3 (z-z_2)}, & z \in (z_2, \; z_2{+}\Delta z_3) \\ 
\cdots \\[0ex]
e^{-\sum\limits_{i=1}^{j-1} \mu_i \Delta z_i} e^{-\mu_j (z-z_{j-1})}, &
z \in (z_{j-1}, \; z_{j-1}{+}\Delta z_j)
\end{dcases} \nonumber
\end{align}
To simplify the presentation, we introduce the sequence $\{ S_j \equiv S(z_j) \}$
and its properties. 
\begin{equation}
\boxed{\quad \begin{aligned}
& S_j \equiv S(z_j) = e^{-\sum\limits_{i=1}^{j} \mu_i \Delta z_i}, \quad j = 1, 2, \ldots, M \\[1ex]
& S_0 = 1, \quad S_M = 0, \quad S_j = S_{j-1} e^{-\mu_j \Delta z_j}
\end{aligned} \quad} \label{Sj_def} 
\end{equation}
$S_j$ is the relative power density passing through layer $j$ and surviving beyond 
interface $z_j$. 
With the notation of $S_j$, we write the $M$ segments of $S(z)$ as a summation. 
\begin{equation}
 S(z) = \sum_{j=1}^M \big(H(z{-}z_{j-1})-H(z{-}z_{j}) \big) S_{j-1} e^{-\mu_j (z-z_{j-1})} 
\label{Sz_sol}
\end{equation}
where $H(z)$ is the Heaviside function. 

\subsection{Skin temperature evolution}
\begin{figure}[!h]
\vskip 0.0cm
\centering
\includegraphics[width=2.75in]{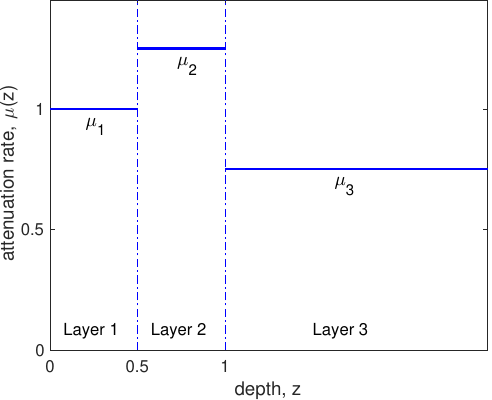} \qquad 
\includegraphics[width=2.75in]{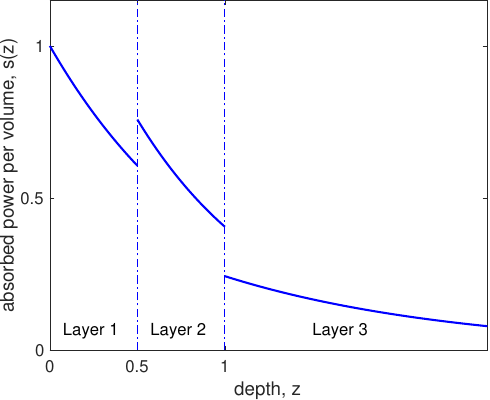}
\vskip -0.0cm
\caption{Left: a schematic profile of 
stratified attenuation rate $\mu(z)$. Right: the corresponding  
absorbed electromagnetic power per volume (the heat source). }
\label{fig_03}
\end{figure}

The heat source $Q(x, y, z)$ is from the attenuation of $P_d^{(a)}(x, y, z) = P_d^{(a)} \rho(x, y) S(z)$. 
\begin{align}
Q(x, y, z) & = -\frac{\partial P_d^{(a)}(x, y, z)}{\partial z } 
= P_d^{(a)} \rho(x, y) s(z), \quad  s(z) \equiv -\frac{d S(z)}{d z} 
\label{Q_sol}
\end{align}
where $S(z)$ is the relative power density surviving beyond depth $z$, given in \eqref{Sz_sol}.
$s(z) \equiv -\frac{d S(z)}{d z}$ is the relative profile of absorbed electromagnetic 
power per unit volume. 
\begin{equation}
\boxed{ \quad s(z) = \sum_{j=1}^M \big(H(z{-}z_{j-1})-H(z{-}z_{j}) \big) \mu_j S_{j-1} 
e^{-\mu_j (z-z_{j-1})} \quad}
\label{sz_sol}
\end{equation} 
Figure \ref{fig_03} shows a schematic profile of 
stratified attenuation rate $\mu(z)$ (left) and the corresponding  
absorbed electromagnetic power per volume (right).

Let $T(x, y, z, t)$ be the skin temperature at position $(x, y, z)$ at time $t$. 
We make a few assumptions in modeling the electromagnetic heating. 
\begin{itemize}
\item 
The skin temperature prior to electromagnetic exposure is uniform in space, 
independent of $(x, y, z)$, 
which is labeled the baseline temperature $T_\text{base}$. 
This assumption gives the initial condition $T(x, y, z, t) \big|_{t=0} = T_\text{base}$. 
This assumption is mainly for simplicity, not essential for the analytical solution. 
A homeostatic initial temperature distribution can be superposed on the analytical solution 
obtained here. 
\item 
The rate of heat loss at the skin surface is neglected. 
The heat loss via convection and thermal radiation is small in comparison 
with the electromagnetic power going into skin. 
This assumption is valid in electromagnetic exposures of
 high intensity and short duration. It leads to the boundary condition 
$\partial_z T(x, y, z, t) \big|_{z=0} = 0$. 
\item 
The heat capacity $\rho_m C_p$ and conductivity $k$ are uniform 
throughout the multi-layer skin 
even though  the attenuation rate $\mu(z)$ has a stratified profile in depth. 
\item 
The last layer of skin extends to infinity: $z_M = +\infty$. 
This assumption is mainly for mathematical convenience. 
Skin is lossy medium and the electromagnetic power decays exponentially while 
propagating in skin. 
The assumption of $z_M = +\infty$ is virtually valid as long as the physical 
thickness of skin tissue under consideration exceeds the depth that the electromagnetic wave 
can practically reach. 
\end{itemize}
With the assumptions above, and the electromagnetic heating source given in \eqref{Q_sol}, 
the skin temperature evolution is governed by conservation of energy. 
\begin{equation}
\begin{dcases}
\underbrace{\rho_m C_p\frac{\partial T}{\partial t}}_{\substack{
\text{rate of change}\\ \text{of heat in skin}}}
= \underbrace{k \Big(\frac{\partial^2T}{\partial x^2}
+\frac{\partial^2T}{\partial y^2}+\frac{\partial^2T}{\partial z^2}
\Big)}_{\substack{
\text{heat flux from}\\ \text{conduction}}} 
+ \underbrace{P_d^{(a)} \rho(x, y) s(z)}_{
\substack{\text{heat source from}\\ \text{absorbed power}}} \\[2ex]
\underbrace{\frac{\partial T(x, y, z, t)}{\partial z} \Big|_{z=0} = 0}_{
\text{no heat loss at skin surface}}, \qquad 
\underbrace{T(x, y, z, 0) = T_\text{base}}_{\text{baseline temperature}}
\end{dcases}
\label{T_eq}
\end{equation}
where $\rho_m$ is the mass density, $ C_p$ the heat capacity,  and $k$ the heat conductivity of the skin. 
\eqref{T_eq} is the governing model of physical temperature in electromagnetic heating. 

\section{Nondimensionalization and Reduction}
\subsection{Physical scales and normalization}
Below we introduce relevant scales for normalizing various physical quantities and 
the resulting nondimensional quantities. 
For a quantity in the model, we use $X_s$ to denote the scale for normalizing the quantity, 
$X$ without $_\text{nondim}$ to denote the nondimensional quantity after normalization. 
For clarity, the original physical quantity, if needed, are denoted by $X_\text{phy}$
\cite{Wang_2020}. We start with the depth scale. For a single layer skin
with attenuation rate $\mu_{phy}$, the characteristic depth of electromagnetic penetrating into skin is 
$1/\mu_{phy}$. For the multi-layer skin with the stratified attenuation rate profile 
$\{\mu_{j, phy}, j = 1, 2, \ldots, M \}$ 
in \eqref{mu_profile}, we set $\mu_s$ for the depth scale to a representative value 
of $\{\mu_{j, phy} \}$, the median, for example. 
\begin{equation}
\begin{dcases}
\mu_s \equiv \text{representative of }\{\mu_{j, phy} \}, \qquad 
\mu_j \equiv \frac{\mu_{j, phy}}{\mu_s} \\[0ex]
z_s \equiv \frac{1}{\mu_s}, \qquad z \equiv \frac{z_\text{phy}}{z_s} \\[0ex]
r_s \equiv (\text{lateral length scale}), \qquad 
(x, y) \equiv \frac{(x_\text{phy}, y_\text{phy})}{r_s} \\[0ex]
t_s \equiv \frac{\rho_m C_p}{k \mu_s^2},\qquad t \equiv \frac{t_\text{phy}}{t_s}
\end{dcases}
\label{scales_nd_qs1} 
\end{equation}
\begin{equation}
\begin{dcases}
T_s\equiv (\text{temperature scale}), \\ 
\qquad T(x, y, z, t) \equiv \frac{T_\text{phy}(x_\text{phy}, y_\text{phy}, z_\text{phy}, t_\text{phy})
-T_\text{base,phy}}{T_s} \\[0ex]
P_s\equiv k \mu_s T_s, \qquad P_\text{d}^{(a)} \equiv\frac{P_\text{d,phy}^{(a)}}{P_s} \\[0ex] 
\rho(x, y) \equiv \rho_\text{phy}(x_\text{phy}, y_\text{phy}) \\[1ex]
S(z) \equiv S_\text{phy}(z_\text{phy}), \qquad S_j \equiv S_{j,phy}
\end{dcases}
\label{scales_nd_qs2} 
\end{equation}
In the above, the lateral length scale $r_s$ is set to the cross-section radius of 
the electromagnetic beam; and the temperature scale $T_s$ is set to 
the physical temperature increase from the baseline to the activation temperature 
of thermal nociceptors in skin. 
All other scales are derived from skin material properties, for example, the time scale 
$t_s \equiv \frac{\rho_m C_p}{k \mu^2}$. 
Both $S_\text{phy}(z_\text{phy})$ and $\rho_\text{phy}(x_\text{phy}, y_\text{phy})$
represent relative power densities; their function values are already nondimensional 
in the physical model; in the nondimensionalization, only their independent variables 
are normalized. 

After normalization, the governing initial boundary value problem (IBVP) \eqref{T_eq} becomes 
\begin{equation}
\begin{dcases}
\displaystyle \frac{\partial T}{\partial t} 
= \varepsilon^2 \Big(\frac{\partial^2T}{\partial x^2}
+\frac{\partial^2T}{\partial y^2}\Big)+\frac{\partial^2T}{\partial z^2}
+P_d^{(a)} \rho(x, y) s(z) \\[2ex]
\displaystyle \frac{\partial T(x, y, z, t)}{\partial z} \bigg|_{z=0}=0, \qquad T(x, y, z, 0) = 0 
\end{dcases}
\label{T_eq2}
\end{equation}
where $\displaystyle \varepsilon \equiv \frac{z_s}{r_s} $ is the depth-to-lateral scale ratio. 
IBVP \eqref{T_eq2} governs the evolution 
of nondimensional temperature $T(x,y,z,t)$. It includes the effect of lateral heat conduction, 
and is valid for arbitrary $\varepsilon$. 
In most MMW exposures, the beam spot radius is much larger than the electromagnetic 
penetration depth. Consequently, the depth-to-lateral scale ratio $\varepsilon$ is small. 
Previously we carried out an asymptotic analysis on \eqref{T_eq2} for small $\varepsilon$
in the case of a single layer skin (i.e., $\mu(z) \equiv \mu$). 
We derived the leading term solution and the two-term asymptotic solution 
that is accurate to $O(\varepsilon^2)$ \cite{WBZ_2020C, Jaime-Yepez_2024}. 

\subsection{Reduction of IBVP \eqref{T_eq2} to a 1D problem}
We extend the study of electromagnetic heating to the case of a multi-layer skin with the stratified 
attenuation rate profile described in \eqref{mu_profile}. 
We derive the analytical solution of \eqref{T_eq2} for $\varepsilon = 0$, which corresponds 
to the leading term asymptotic solution for small $\varepsilon$. 
When the lateral heat conduction is neglected ($\varepsilon = 0$),  \eqref{T_eq2} becomes 
\begin{equation}
\begin{dcases}
\displaystyle \frac{\partial T}{\partial t} 
= \frac{\partial^2T}{\partial z^2}+P_d^{(a)} \rho(x, y) s(z) \\[2ex]
\displaystyle \frac{\partial T(x, y, z, t)}{\partial z} \bigg|_{z=0}=0, \qquad T(x, y, z, 0) = 0 
\end{dcases}
\label{T_eq2_ep0}
\end{equation}
In IBVP \eqref{T_eq2_ep0}, $(x, y)$ are no longer differential variables; they are parameters. 
In addition, i) the heat source $\rho(x, y) s(z)$ have separate dependences 
on $(x, y)$ and on $z$, ii) \eqref{T_eq2_ep0} is linear with homogeneous initial 
and boundary conditions. It follows from these facts that the solution $T(x, y, z, t)$
of \eqref{T_eq2_ep0} has the same lateral spatial profile as the heat source. 
\begin{equation}
\boxed{\quad T(x, y, z, t) = P_d^{(a)} \rho(x, y) T^{(1)}(z, t) \quad } \label{sol_0}
\end{equation}
where $T^{(1)}(z, t)$ is the nondimensional temperature at the beam center 
per unit power density. $T^{(1)}(z, t)$ is a function of $(z, t)$ only and is governed by a 1D problem. 
\begin{equation}
\begin{dcases}
\displaystyle \frac{\partial T^{(1)}}{\partial t} 
= \frac{\partial^2 T^{(1)}}{\partial z^2}+s(z), \quad z \in(0, +\infty) \\[2ex]
\displaystyle \frac{\partial T^{(1)}(z, t)}{\partial z} \bigg|_{z=0}=0, \quad T^{(1)}(z, 0) = 0 
\end{dcases}
\label{T_eq3}
\end{equation}
where the heat source $s(z)$ consists of multiple segments as given in \eqref{sz_sol}. 

\subsection{Reduction of IBVP \eqref{T_eq3} to a parameter-free IVP}
We first consider the initial value problem (IVP) below related to IBVP \eqref{T_eq3}.
\begin{equation}
\begin{dcases}
\displaystyle \frac{\partial T^{(2)}}{\partial t} 
= \frac{\partial^2T^{(2)}}{\partial z^2}+ s(z), \quad z \in(-\infty, +\infty) \\[1ex]
\displaystyle T^{(2)}(z, 0) = 0 
\end{dcases}
\label{sys_2}
\end{equation}
Note that the heat source $s(z)$ given in \eqref{sz_sol} already satisfies $s(z) = H(z) s(z)$. 
The zero-flux boundary condition in \eqref{T_eq3} is taken care of with an even image reflection 
on \eqref{sys_2}. Solutions $T^{(1)}$  of \eqref{T_eq3} and solution $T^{(2)}$ of \eqref{sys_2} 
are related by 
\begin{equation}
\boxed{ \quad T^{(1)}(z, t) = T^{(2)}(z, t) + T^{(2)}(-z, t), \quad z > 0 \quad}
\label{sol_1}
\end{equation} 
With \eqref{sol_0} and \eqref{sol_1}, we see that to find the 3D temperature distribution 
$T(x, y, z, t)$, we only need to solve for $T^{(2)}$ in \eqref{sys_2}. 
The heat source $s(z) $ of \eqref{sys_2}, as detailed in \eqref{sz_sol}, has multiple segments. 
To simplify the presentation, we introduce function $f(z; \lambda, z_c)$. 
\begin{equation}
f(z; \lambda, z_c) \equiv H(z-z_c) e^{-\lambda (z-z_c)}
\label{fz_def}
\end{equation}
We express the heat source $s(z) $ in terms of $f(z; \lambda, z_c) $. 
\begin{align}
& s(z) = \sum_{j=1}^M \mu_j \Big( S_{j-1}H(z{-}z_{j-1}) e^{-\mu_j (z-z_{j-1})}
- S_{j-1}e^{-\mu_j \Delta z_j} H(z{-}z_j) e^{-\mu_j (z-z_j)} \Big) \nonumber \\[0ex]
& \qquad = \sum_{j=1}^M \mu_j \Big(S_{j-1} f(z; \mu_j, z_{j-1}) 
- S_j f(z; \mu_j, z_j) \Big) \label{sz_sol2} 
\end{align}
$f(z; \lambda, z_c)$ is the exponential function, horizontally scaled, shifted and truncated. 
For various combinations of $(\lambda, z_c)$, functions $\{ f(z; \lambda, z_c) \}$ 
serve as the building blocks for representing $s(z)$, 
the profile of absorbed electromagnetic power per volume in a multi-layer skin. 
Let $U(z, t; \lambda, z_c)$ be the solution of the IVP with $f(z; \lambda, z_c)$ as the source term. 
\begin{equation}
\begin{dcases}
\displaystyle \frac{\partial U}{\partial t} 
= \frac{\partial^2 U}{\partial z^2}+ \underbrace{\;\; f(z; \lambda, z_c) \;\; }_{
\equiv H(z{-}z_c) e^{-\lambda (z{-}z_c)}}, \quad z \in(-\infty, +\infty) \\[1ex]
\displaystyle U(z, 0) = 0 
\end{dcases}
\label{sys_3}
\end{equation}
IVP \eqref{sys_2} is linear with homogeneous initial condition. 
The source term $s(z)$ of \eqref{sys_2} is a linear combination of 
functions $\{f(z; \lambda, z_c) \}$. It follows that solution $T^{(2)}$ of \eqref{sys_2} 
is a linear combination of solutions $U(z, t; \lambda, z_c)$.
\begin{equation}
T^{(2)}(z, t) = \sum_{j=1}^M \mu_j \Big(S_{j-1} U(z, t; \mu_j, z_{j-1}) 
- S_j U(z, t; \mu_j, z_j) \Big) 
\label{sol_3}
\end{equation}
Thus, we only need to solve for $U(z, t; \lambda, z_c)$ in \eqref{sys_3}. 
We reduce \eqref{sys_3} further to a parameter-free problem 
with the change of variables below. 
\begin{equation}
 z_\text{new} = \lambda (z-z_c), \quad t_\text{new} = \lambda ^2 t, \quad 
 V(z_\text{new},  t_\text{new}) =  \lambda^2 U(z, t)
 \label{change_vars}
\end{equation}
For simplicity, in the resulting IVP, $ V(z_\text{new},  t_\text{new}) $ is abbreviated to $V(z, t)$. 
\begin{equation}
\begin{dcases}
\displaystyle \frac{\partial V}{\partial t} 
= \frac{\partial^2 V}{\partial z^2}+ H(z) e^{-z}, \quad z \in(-\infty, +\infty) \\[1ex]
\displaystyle V(z, 0) = 0 
\end{dcases}
\label{sys_4}
\end{equation}
Note that IVP \eqref{sys_4} and its solution $V(z, t)$ are both parameter-free. 
Solutions $U(z, t; \lambda, z_c)$ of \eqref{sys_3} and solution $V(z, t)$ of \eqref{sys_4} are related by 
\begin{equation}
U(z, t; \lambda, z_c) = \frac{1}{\lambda^2} V\big(\lambda(z-z_c), \lambda^2 t \big) 
\label{sol_4}
\end{equation} 
We merge \eqref{sol_3} and \eqref{sol_4} to express $T^{(2)}(z, t)$ in terms of 
the parameter-free $V(z, t)$. 
\begin{equation} 
\boxed{\;\; T^{(2)}(z, t) = \sum_{j=1}^M \frac{1 }{\mu_j}\Big(
S_{j-1} V\big(\mu_j(z-z_{j-1}), \mu_j^2 t \big) 
- S_{j} V\big(\mu_j(z-z_j), \mu_j^2 t \big) \Big) \;\;}
\label{sol_3B}
\end{equation}
Therefore, to find $T^{(2)}(z, t)$, we only need to solve for $V(z, t)$ in IVP \eqref{sys_4}.

\section{Solution of the Parameter-Free IVP \eqref{sys_4}}
\subsection{Reduction to an IVP with zero source}
\eqref{sys_4} has zero initial value and a nonhomogeneous source term $H(z) e^{-z}$. 
We apply Duhamel's principle to solve it. We consider the associated IVP with zero source 
and a nonhomogeneous initial condition reflecting the impulse effect of $H(z) e^{-z}$. 
\begin{equation}
\begin{dcases}
\displaystyle \frac{\partial G}{\partial t} 
= \frac{\partial^2 G}{\partial z^2}, \quad z \in(-\infty, +\infty) \\[1ex]
\displaystyle G(z, 0) =  H(z) e^{-z} 
\end{dcases}
\label{sys_G}
\end{equation}
Solution $G(z, t)$ is given by the convolution of $H(z) e^{-z}$ 
with the fundamental solution. 
\begin{align*}
& G(z, t) = \int_{-\infty}^{+\infty} H(z') e^{-z'} \frac{1}{\sqrt{4 \pi t}} e^{\frac{-(z'-z)^2}{4t}} d z' \\[1ex]
& \qquad = \frac{1}{\sqrt{4 \pi t}} \int_{0}^{+\infty} e^{\frac{-(z'^2-2z' (z-2t) + z^2)}{4t}} d z' \\[1ex]
& \qquad \xrightarrow{\; \xi = z' - (z-2t) \; } 
\frac{e^{-z+t}}{\sqrt{4 \pi t}} \int_{-(z-2t)}^{+\infty} e^{\frac{-\xi^2 }{4t}} d \xi \\[1ex]
& \qquad \xrightarrow{\; \eta = \xi/\sqrt{4t} \; } 
\frac{e^{-z+t}}{\sqrt{\pi}} \int_{\frac{-z+2t}{\sqrt{4t}}}^{+\infty} e^{-\eta^2 } d \eta 
\nonumber
\end{align*}
\begin{equation}
G(z, t) = \frac{e^{-z+t}}{2} \text{erfc}(\frac{-z+2t}{\sqrt{4t}}) 
\label{G_sol}
\end{equation}
where $\text{erfc}(x)$ is the complementary error function defined as 
\begin{equation}
\text{erfc}(x) \equiv \frac{2}{\sqrt{\pi}} \int_x^{+\infty} e^{-\eta^2} d\eta 
\label{erfc_def}
\end{equation}
The asymptotic behaviors of $\text{erfc}(x)$ and $G(z, t)$ are 
\begin{equation}
\begin{dcases}
\lim_{x \rightarrow -\infty} \text{erfc}(x) = 2, \qquad 
	\lim_{x \rightarrow +\infty} \text{erfc}(x) = 0 \\[1ex]
e^{x^2} \text{erfc}(x) \sim \frac{1}{x \sqrt{\pi}} \quad \text{as $x \rightarrow +\infty$} \\[1ex]
G(z, t) = \frac{e^{-z+t}}{2} \text{erfc}(\frac{-z+2t}{\sqrt{4t}}) \sim e^{-z+t}
\quad \text{as $z \rightarrow +\infty$} \\[1ex]
G(z, t) = \frac{1}{2} e^{\frac{-z^2}{4t}} e^{(\frac{-z+2t}{\sqrt{4t}})^2} \text{erfc}(\frac{-z+2t}{\sqrt{4t}}) 
\sim \sqrt{\frac{t}{\pi}}\, \frac{1}{|z|} e^{\frac{-z^2}{4t}} \quad \text{as $z \rightarrow -\infty$}
\end{dcases}
\label{asymp_1}
\end{equation}
Solution $V(z, t)$ of \eqref{sys_4} and solution $G(z, t)$ of \eqref{sys_G} 
are related by Duhamel's Principle. 
\begin{equation}
V(z, t) = \int_0^t G(z, t-s) ds = \int_0^t \frac{e^{-z+s}}{2} \text{erfc}(\frac{-z+2s}{\sqrt{4s}}) ds
\label{Duhamel}
\end{equation}

\subsection{Evaluate the integral in Duhamel's Principle}
To prepare for integration by parts, we fix $z$ and derive several results. 
\begin{align}
& \frac{d \big[\text{erfc}(x) \big]}{dx} = \frac{-2}{\sqrt{\pi}} e^{-x^2}, \qquad 
e^{-z+s} e^{-(\frac{-z+2s}{\sqrt{4s}})^2} = e^{\frac{-z^2}{4s}}
\label{Res_1} \\[1ex]
& d \Big[ \text{erfc}(\frac{-z+2s}{\sqrt{4s}})\Big] = \frac{-2}{\sqrt{\pi}}
e^{-(\frac{-z+2s}{\sqrt{4s}})^2} \Big(\frac{z}{2s}+1\Big) d(\sqrt{s})  
\label{Res_2} \\[1ex]
& d \Big[\text{erfc}(\frac{z}{\sqrt{4s}})\Big] = \frac{2}{\sqrt{\pi}}
e^{\frac{-z^2}{4s}}\frac{z}{2s} d(\sqrt{s}) \label{Res_4} \\[1ex]
& d\big[ \sqrt{s}\, e^{\frac{-z^2}{4s}}\big] = e^{\frac{-z^2}{4s}}\Big(1+ \frac{z^2}{2s}\Big) d(\sqrt{s}) 
= e^{\frac{-z^2}{4s}}d(\sqrt{s})+ z \frac{\sqrt{\pi}}{2} d \Big[\text{erfc}(\frac{z}{\sqrt{4s}})\Big]
\label{Res_5}
\end{align}
In the above \eqref{Res_4} is used in \eqref{Res_5}.
Note that $\text{erfc}(\frac{-z+2t}{\sqrt{4t}})$ has singularity at $(z = 0, \, t = 0)$. 
\[ \text{erfc}(\frac{-z+2t}{\sqrt{4t}}) \bigg|_{t = 0_+} = 
\begin{dcases}
2, & z > 0 \\ 
0, & z < 0
\end{dcases} \]
Because of this singularity, we discuss $z > 0$ and $z < 0$ separately. 
\begin{itemize}
\item
For $z > 0$, integration by parts in \eqref{Duhamel} gives 
\begin{equation}
\begin{aligned}
& V(z, t) = \frac{e^{-z+s}}{2} \text{erfc}(\frac{-z+2s}{\sqrt{4s}}) \Big|_{s=0_+}^{s=t} 
-\underbrace{\int_0^t \frac{e^{-z+s}}{2} d \Big[ \text{erfc}(\frac{-z+2s}{\sqrt{4s}}) \Big]}_{
\text{using \eqref{Res_2} and \eqref{Res_1}}} \\
& \quad = \frac{e^{-z+t}}{2} \text{erfc}(\frac{-z+2t}{\sqrt{4t}}) - e^{-z}
+ \frac{1}{\sqrt{\pi}} \int_0^t e^{\frac{-z^2}{4s}} 
\Big(\frac{z}{2s}+1\Big) d(\sqrt{s})
\end{aligned}
\nonumber
\end{equation}
We evaluate the two parts in the integral above using respectively \eqref{Res_4} and \eqref{Res_5}. 
\begin{align*}
& \frac{1}{\sqrt{\pi}} \int_0^t e^{\frac{-z^2}{4s}} \frac{z}{2s} d(\sqrt{s}) 
= \frac{1}{2}\int_0^t d \Big[\text{erfc}(\frac{z}{\sqrt{4s}})\Big] 
= \frac{1}{2}\text{erfc}(\frac{z}{\sqrt{4t}}), \quad z > 0 \\[2ex]
& \frac{1}{\sqrt{\pi}} \int_0^t e^{\frac{-z^2}{4s}} d(\sqrt{s}) 
= \frac{1}{\sqrt{\pi}} \int_0^t  d\big[ \sqrt{s}\, e^{\frac{-z^2}{4s}}\big] 
- \frac{z}{2} \int_0^t  d \Big[\text{erfc}(\frac{z}{\sqrt{4s}})\Big] 
\\[0ex]
& \qquad \qquad = \frac{1}{\sqrt{\pi}} \sqrt{t}\, e^{\frac{-z^2}{4t}}- \frac{z}{2}\text{erfc}(\frac{z}{\sqrt{4t}}),
\quad z > 0 
\end{align*}
For $z > 0$, the expression of $V(z, t)$ is 
\begin{equation}
\begin{aligned}
& V(z, t) = \frac{e^{-z+t}}{2} \text{erfc}(\frac{-z+2t}{\sqrt{4t}}) 
+ \sqrt{\frac{t}{\pi}}\, e^{\frac{-z^2}{4t}}-\frac{1+z}{2}\text{erfc}(\frac{z}{\sqrt{4t}}) \\
& \qquad \qquad  + \Big(\text{erfc}(\frac{z}{\sqrt{4t}})- e^{-z}\Big), \qquad z > 0
\end{aligned}
\label{Vz_posi}
\end{equation}
\item 
For $z < 0$, integration by parts in \eqref{Duhamel} yields 
\begin{equation}
\begin{aligned}
& V(z, t) = \frac{e^{-z+t}}{2} \text{erfc}(\frac{-z+2t}{\sqrt{4t}}) 
+ \frac{1}{\sqrt{\pi}} \int_0^t e^{\frac{-z^2}{4s}} 
\Big(\frac{z}{2s}+1\Big) d(\sqrt{s})
\end{aligned}
\nonumber
\end{equation}
We evaluate the two parts in the integral above using respectively \eqref{Res_4} and \eqref{Res_5}. 
\begin{align*}
& \frac{1}{\sqrt{\pi}} \int_0^t e^{\frac{-z^2}{4s}} \frac{z}{2s} d(\sqrt{s}) 
= \frac{1}{2}\int_0^t d \Big[\text{erfc}(\frac{z}{\sqrt{4s}})\Big] \\[0ex]
& \qquad \qquad = \frac{1}{2}\text{erfc}(\frac{z}{\sqrt{4t}}) -1 
= \frac{-1}{2}\text{erfc}(\frac{-z}{\sqrt{4t}}), \quad z < 0 \\[2ex]
& \frac{1}{\sqrt{\pi}} \int_0^t e^{\frac{-z^2}{4s}} d(\sqrt{s}) 
= \frac{1}{\sqrt{\pi}} \int_0^t  d\big[ \sqrt{s}\, e^{\frac{-z^2}{4s}}\big] 
- \frac{z}{2} \int_0^t  d \Big[\text{erfc}(\frac{z}{\sqrt{4s}})\Big] \\[0ex]
& \qquad \qquad = \frac{1}{\sqrt{\pi}} \sqrt{t}\, e^{\frac{-z^2}{4t}}- \frac{z}{2}
\Big( \text{erfc}(\frac{z}{\sqrt{4t}})-2 \Big) \\[0ex]
& \qquad \qquad = \frac{1}{\sqrt{\pi}} \sqrt{t}\, e^{\frac{-z^2}{4t}}- \frac{(-z)}{2}
\text{erfc}(\frac{-z}{\sqrt{4t}}), \quad z < 0  
\end{align*}
For $z < 0$, the expression of $V(z, t)$ is 
\begin{equation}
\begin{aligned}
& V(z, t) = \frac{e^{-z+t}}{2} \text{erfc}(\frac{-z+2t}{\sqrt{4t}}) 
+ \sqrt{\frac{t}{\pi}}\, e^{\frac{-z^2}{4t}}-\frac{1+(-z)}{2}\text{erfc}(\frac{(-z)}{\sqrt{4t}}),
\quad z < 0
\end{aligned}
\label{Vz_nega}
\end{equation}
\end{itemize}
We synthesize \eqref{Vz_posi} for $z > 0$ and \eqref{Vz_nega} for $z < 0$ into a unified expression. 
\begin{equation}
\boxed{\quad 
\begin{aligned}
& V(z, t) = \frac{e^{-z+t}}{2} \text{erfc}(\frac{-z+2t}{\sqrt{4t}}) 
+ \sqrt{\frac{t}{\pi}}\, e^{\frac{-z^2}{4t}}-\frac{1+|z|}{2}\text{erfc}(\frac{|z|)}{\sqrt{4t}}) \\
& \qquad \qquad  +
H(z) \big(\text{erfc}(\frac{|z|}{\sqrt{4t}})- e^{-|z|}\big)
\end{aligned} \quad }
\label{V_unified}
\end{equation}
In \eqref{V_unified}, every term is well behaved as $z \rightarrow \pm \infty$. The last term has 
two segments, which are necessary for accommodating that the electromagnetic heating source 
is active only in $z > 0$. Despite the two segments, function $V(z, t)$ is continuous at $z = 0$. 

\subsection{Roadmap of solving for $T(x, y, z, t)$ in \eqref{T_eq2_ep0}}
Now we summarize the process of solving for the 3D skin temperature distribution 
$T(x, y, z, t)$. In electromagnetic heating of a multi-layer 
skin with layer-specific attenuation rates, when the effect of lateral heat conduction 
is neglected $(\varepsilon = 0)$, the solution of nondiemnsional governing system 
\eqref{T_eq2_ep0} is completely described by two groups of inputs: 
\begin{enumerate}
\item the geometric and material parameters of skin layers $\big\{(z_j)_{j=0}^M, \; (\mu_j)_{j=1}^M \big\}$ and
\item the specifications of electromagnetic beam $\big\{P_d^{(a)}, \; \rho(x, y) \big\}$
\end{enumerate}
Skin layer parameters $\big\{(z_j)_{j=0}^M, \; (\mu_j)_{j=1}^M \big\}$ are hidden 
in the stratified profile of relative heat source $s(z)$ as detailed in \eqref{sz_sol}. 
The overall process of solving $T(x, y, z, t)$ is 
\begin{equation}
\boxed{\;\; \big\{(z_j)_{j=0}^M, \; (\mu_j)_{j=1}^M, \; P_d^{(a)}, \; \rho(x, y) \big\} 
\xrightarrow{\;\; \text{\eqref{Sj_def}, \eqref{V_unified},  
\eqref{sol_3B}, \eqref{sol_1}, \eqref{sol_0}}\;\; } \; T(x, y, z, t) \;\;}
\label{T_sol}
\end{equation}
It consists of five steps. 
\begin{equation}
\begin{dcases}
\text{parameters } \big\{(z_j)_{j=0}^M, \; (\mu_j)_{j=1}^M \big\}  
\xrightarrow{\quad \text{\eqref{Sj_def} }\quad } 
\; (S_j)_{j=0}^M \\[1ex] 
\text{paramneter-free} \xrightarrow{\quad \text{\eqref{V_unified}}\quad } 
\; V(z, t) \\[1ex] 
\text{parameters } \big\{(z_j)_{j=0}^M, \; (\mu_j)_{j=1}^M, \; (S_j)_{j=0}^M \big\}
+V(z, t) \xrightarrow{\quad \text{\eqref{sol_3B}}\quad } 
\; T^{(2)}(z, t) \\[1ex] 
T^{(2)}(z, t) \xrightarrow{\quad \text{\eqref{sol_1}}\quad } \; T^{(1)}(z, t)  \\[1ex]
\text{parameters } \big\{P_d^{(a)} \big\}+\rho(x, y)+ T^{(1)}(z, t) 
\xrightarrow{\quad \text{\eqref{sol_0}}\quad } \; T(x, y, z, t)
\end{dcases}
\label{T_sol_detail}
\end{equation}

\subsection{Numerical verification}
\begin{figure}[!h]
\vskip 0.0cm
\centering
\includegraphics[width=2.75in]{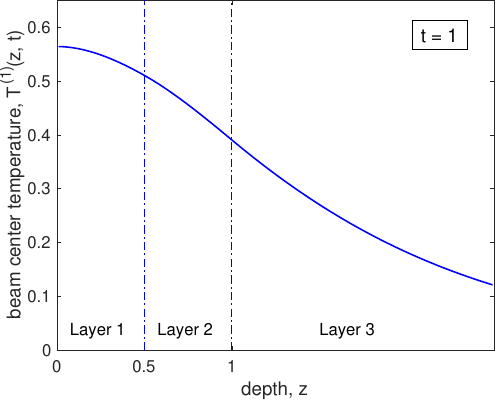} \qquad 
\includegraphics[width=2.75in]{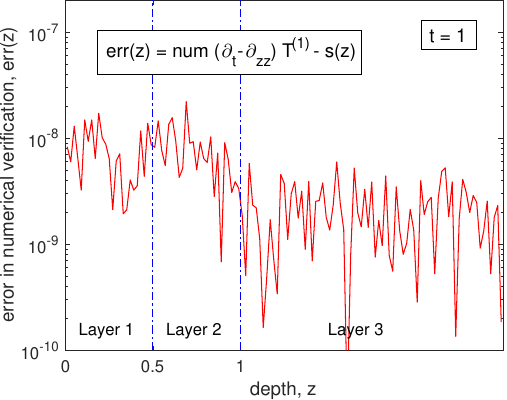}
\vskip -0.0cm
\caption{Results of $T^{(1)}$ from \eqref{T_sol_detail}. 
Left: beam center temperature per unit power density. 
Right: error in numerical verification of equation \eqref{T_eq3} on $T^{(1)}$. }
\label{fig_04}
\end{figure}

We use a numerical approach to verify the analytical expressions derived above. 
We examine $T^{(1)}(z, t)$ from \eqref{T_sol_detail}, which is the nondimensional 
beam center temperature per unit power density. 
$T^{(1)}(z, t)$ is governed by equation \eqref{T_eq3}. 
We write a code based on the analytical expression from \eqref{T_sol_detail} 
to compute function values of $T^{(1)}(z, t)$. 
We use numerical differentiation to approximate 
$\big( \partial_t - \partial_{zz} \big) T^{(1)}$ in \eqref{T_eq3}. 
To verify that the $T^{(1)}(z, t)$ expression 
derived above is correct, we check the residual error in \eqref{T_eq3}.  
\begin{equation}
\text{err}(z) \equiv \big[ \text{numerical } \big( \partial_t - \partial_{zz} \big) T^{(1)} \big] -s(z) 
\label{err_def}
\end{equation}
In our numerical verification, we use the skin layer parameters below. 
\begin{align*}
M = 3, \qquad & z_1 =0.5, \qquad z_2 = 1, \qquad z_3 = +\infty \\
& \mu_1 = 1, \qquad \mu_2 =1.25, \qquad \mu_3 = 0.75  
\end{align*}
In Figure \ref{fig_04}, the left panel shows $T^{(1)}(z, t) \text{ vs } z$ at $t = 1$; 
the right panel plots $\text{err}(z)$, the residual error in the numerical verification. 
The residual error confirms that the analytical expression of $T^{(1)}(z, t)$ obtained above 
satisfies equation \eqref{T_eq3} to the numerical accuracy of central difference 
for the second derivative in IEEE double precision.  

\section{Concluding Remarks}
We studied the electromagnetic heating of a skin tissue consisting of 
multiple layers with layer specific attenuation rates of electromagnetic power. 
We focused on the case where the incident beam is perpendicular to the flat skin surface
and the beam spot size is much larger than the electromagnetic penetration depth into the skin. 
In this case, the effect of lateral heat conduction is negligible at the leading term. 
We described the model of electromagnetic power absorption in the multi-layer skin 
where the absorbed power becomes the heat source. 
The resulting skin temperature evolution is driven by 
the absorbed electromagnetic power. 
We nondimensionalized the governing model and expressed the solution of 
the nondimensional system as a combination of building blocks, reflecting 
the effects of stratified electromagnetic absorption in multiple skin layers. 
Mathematically, all building blocks used are variants in the same family. 
We used scaling and shifting to reduce the governing equations of building blocks 
to a parameter-free initial value problem of the heat equation with the standard exponentially 
decaying heat source. We derived the analytical solution of the associated initial value problem
with zero heat source. We applied Duhamel's principle and integrated the analytical solution 
of the associated problem in time to obtain a close-form analytical solution of 
the parameter-free problem. Then we traced back the pathway of analytical reduction involved 
in arriving at the final parameter-free problem. Based on the reduction pathway, we constructed
the roadmap of solving the 3D skin temperature distribution $T(x, y, z, t)$ 
from two groups of inputs: 
i) geometric and material parameters of skin layers, and 
ii) specifications of electromagnetic beam 
Finally, we used central difference to approximate the derivatives in the governing equation
and verified numerically the validity of the analytical solution by checking the residual 
error in the governing equation.



\end{document}